\documentclass[12pt,preprint]{aastex}

\newcommand{\kms}{\hbox{km s$^{-1}$}}

\newcommand{\mdotyr}{\hbox{$M_\odot$ yr$^{-1}$}}

\shorttitle{Tiny Accretion Rates}
\shortauthors{Herczeg et al.}

\begin{document}

\title{Measuring tiny mass accretion rates onto young brown dwarfs}



\author{Gregory J. Herczeg\altaffilmark{1}, Kelle L. Cruz\altaffilmark{2}, \& Lynne A. Hillenbrand}
      \affil{Astronomy Department, California Institute of Technology, Pasadena, CA 91125, USA}
      \altaffiltext{1}{Current address: Max-Planck-Institut fur extraterrestriche Physik, Postfach 1312, 85741 Garching,
      Germany.  gregoryh@mpe.mpg.de}
      \altaffiltext{2}{Spitzer Postdoctoral Fellow}

\begin{abstract}
We present low-resolution Keck I/LRIS spectra spanning from 3200--9000 \AA\ of nine young brown dwarfs and three low-mass stars in the TW Hya Association and in Upper Sco.   The optical spectral types of the brown dwarfs range from M5.5-M8.75, though two have near-IR spectral types of early L-dwarfs.  We report new accretion rates derived from excess Balmer continuum emission for the low-mass stars TW Hya and Hen 3-600A and the brown dwarfs 2MASS J12073347-3932540,  UScoCTIO 128, SSSPM J1102-3431, UScoJ160606.29-233513.3, DENIS-P J160603.9-205644, and Oph J162225-240515B, and upper limits on accretion for the low-mass star Hen 3-600B and the brown dwarfs UScoCTIO 112, Oph J162225-240515A, and USco J160723.82-221102.0.  For the six brown dwarfs in our sample that are faintest at short wavelengths, the accretion luminosity or upper limit is measurable only when the image is binned over large wavelength intervals.  This method extends our sensivity to accretion rate down to $\sim 10^{-13}$ $\mdotyr$ for brown dwarfs. Since the ability to measure an accretion rate from excess Balmer continuum emission depends
on the contrast between excess continuum emission and the underlying
photosphere, for objects with earlier spectral types the upper limit on accretion
rate is much higher.   Absolute uncertainties in our accretion rate measurements of $\sim 3-5$ include uncertainty in accretion models, brown dwarf masses, and distance.  The accretion rate of $2\times10^{-12}$ $M_\odot$ yr$^{-1}$ onto 2MASS J12073347-3932540 is within 15\% of two previous measurements, despite large changes in the H$\alpha$ flux.
\end{abstract}
 

%
%


\section{INTRODUCTION}
That young brown dwarfs share many characteristics with T Tauri stars is strong evidence that the formation and early evolution of stars and brown dwarfs is similar.  However, the smaller masses of brown dwarfs introduce physical differences that alter the early evolution of the star+disk system, including the prospects for planet formation.  Protoplanetary disks around very low-mass stars and brown dwarfs survive longer than around solar-mass stars \citep{Car06,Dah07,Sch07}, which may relate to the $\dot{M}\propto M^2$ relationship that describes the average accretion rate for a population of objects with a large range of masses \citep[e.g.,][]{Muz05,Moh05,Nat06,Ale06,Har06,Til08}.  
Since accretion requires the presence of a disk, the identification of accretors contributes to estimating the fraction of young stars and brown dwarfs with disks.  Accretion rate measurements lead to estimates for the timescale of viscous dissipation of the protoplanetary disk.  Accretion processes also supply the far-ultraviolet emission \citep{Joh00} that evaporates the outer disk \citep{Gor08} and the ionizing emission that could evaporate the inner disk as accretion slows \citep{Cla01,Ale06a}.

The proximity and age of the $\sim 10$ Myr old TW Hya Association \citep[TWA, d=30--100 pc][]{del89,Gre92,Web99,Mam05} and the $\sim 5$ Myr old Upper Sco Association \citep[USco, d=145 pc,][]{deG89,Wal94,deZ99,Pre02} make them important regions for studying disk dissipation processes around stars and brown dwarfs.  Two of the five known brown dwarfs in the TWA, 2MASS J12073347-3932540 and SSSPM J1102-3431, still retain their disks, though of the two only 2MASS J12073347-3932540 is known as an accretor \citep{Giz02,Moh03,Mor08,Ria08}.  In USco,  \citet{Car06} and \citet{Sch07} find that the fraction of disks around M-dwarfs is larger than that around G and K-type stars, indicating a possible mass-dependence in disk dissipation.  Recently,  \citet{Lod08} used UKIDSS \citep{Law07} to identify 22 brown dwarfs in USco, with 17 classified as L0-L2 dwarfs based on low-resolution near-IR spectroscopy.  If the near-IR spectral types accurately reflect the true mass of the object, an analysis of disks and accretion for the UKIDSS sample would extend our understanding of disk survival to the upper planetary-mass regime.

Accretion can be identified by line equivalent widths and profile shapes \citep[e.g.,][]{Whi03,Muz03,Nat04,Moh05}, but is most accurately measured from excess Balmer continuum at $<3646$ \AA\ \citep{Val93,Gul98}.  For stars with masses 0.5--1.0 $M_\odot$,  accretion continuum measurements typically lead to mass accretion rates $\dot{M}\sim 10^{-9}-10^{-7}$ ${M_\odot}$ yr$^{-1}$.   Upper limits are determined by the ability to measure the contrast between the star and hot photospheric emission, with 
smaller upper limits for cooler targets, which have weaker and redder photospheres.    \citet{Her08} measured the Balmer continuum emission from very low-mass stars and brown dwarfs in Taurus and from 2MASS J12073347-3932540 in the TWA, with accretion rates as low as $2\times10^{-12}$ $M_\odot$ yr$^{-1}$.   Measuring even lower mass accretion rates onto very late M-dwarfs is limited primarily by telescopic sensitivity.

In this paper, we develop a method to extract very low-resolution spectra of the photospheric and accretion continuum from {\it Keck I}/LRIS optical spectra of faint sources.  This technique is applied to a small sample of six brown dwarfs in the TWA and USco to measure accretion rates down to $10^{-13}$ $M_\odot$ yr$^{-1}$, an order of magnitude lower than previous measurements.  We also use traditional spectral extraction techniques to measure accretion or upper limits onto two other brown dwarfs and three stars, and weak emission lines to measure an upper limit onto another brown dwarf.  Our observations and data reduction are described in \S 2.  In \S 3 we describe the physical properties of our sample, identify accretion based on excess blue and UV continuum emission, and use the excess to measure mass accretion rates.  In \S 4 we discuss the implications our accretion rates have for sensitivity to tiny accretion rates and the differences in optical and near-IR spectral types for two USco members.

\section{OBSERVATIONS}
We used LRIS \citep{Oke95,McC98} on {\it Keck I}  to obtain low-resolution optical spectra of nine brown dwarfs in the TWA and Upper Sco and three low-mass stars in the TWA on 28 May 2008 (see observation log in Table 1).  Conditions were clear and stable, with seeing ranging from $0\farcs6-0\farcs8$.  The light was split into separate red and blue beams with the D560 dichroic.  The 400/8500 grating disperses the light in the red channel to yield spectra from 5500--9300 \AA\ on a 2048x2048 CCD with $0\farcs21$ pixels.  The 400/3400 grism in the blue channel disperses the light to yield spectra from 3000--5600 \AA\ on a 2048x4096 CCD with $0\farcs135$ pixels.   Each target was observed with a $175^{\prime\prime}\times1\farcs0$ slit.  The spectral resolution is $\sim1400$ on the red side and $\sim 700$ on the blue side.  

The position angle of the slit was aligned to the parallactic angle for most observations.  During our observations of the visual pairs USco J160723.82-221102.0, USco 160603.75-221930.0, Oph J162225-240515, and Hen 3-600, the slit angle was aligned to include both objects.  The Atmospheric Dispersion Corrector on LRIS \citep{Phi06} corrects the atmospheric dispersion to better than $<0\farcs1$ for observations obtained within $60^\circ$ of the zenith.  The spectrophotometric standards LTT3864, EG 274, and LTT7987 were observed throughout the night to provide flux calibration and extinction correction.   Absolute fluxes are accurate to $\sim 15$\% and relative fluxes across the 3200--9300 \AA\ region are accurate to $\sim 5$\%.  Our observation of USco J160723.82-221102.0 was obtained at $\sim2.4$ airmasses at a position angle $55^\circ$ from the parallactic angle.  The atmospheric dispersion for the observation of USco J160723.82-221102.0 is expected to be larger than $0\farcs5$ at $<5000$ \AA, which prevents us from placing meaningful upper limits on the undetected blue emission from the brown dwarf.

The data were bias- and dark-subtracted, and then flat-fielded with a combination of dome flats and a halogen lamp.  The counts from the two components of the close binary Hen 3-600 were extracted after fitting and subtracting the counts in the image that are attributed to the other source, based on the line-spread function at each spectral position on the detector.

The continuum emission at 3200--4500 \AA\ from USco J160603.75-221930.0, SSSPM J1102-3431, DENIS-P J160603.9-205644, UScoCTIO 112, Oph J162225-240515A, and Oph J162225-240515B is too faint for simple long-slit extraction.  Very low-resolution spectra of the continuum emission are obtained for these sources by binning the image along the dispersion direction.  The counts are then extracted by fitting a Gaussian profile to the counts profile in the cross-dispersion direction.  The $1\sigma$ error bars are estimated from the uncertainty in the background subtraction.  This method significantly reduces the noise introduced when calculating the background subtraction independently at each pixel in the dispersion direction.  

The number and width of spectral bins is individually determined for each target and is informed by the expected shape of the spectrum as seen from brighter accretors and non-accretors and by requiring detections or stringent upper limits in each bin.  The Balmer line blend region from 3646--3700 \AA\ is avoided for  SSSPM J1102-3431, USco 160603.75-221930.0, DENIS-P J160603.9-205644 and UScoCTIO 112 but is included for Oph J162225-240515A and Oph J162225-240515B to increase the signal-to-noise.  This inclusion is reasonable because the flux in the 3646--3700 \AA\ region is very similar to that in the 3600-3646 \AA\ region for all previously-observed M-dwarf accretors and non-accretors \citep{Her08}.  If we restrict this data point to the 3500--3646 \AA\ region, the flux is detected with only marginal significance.

The atmospheric transmission curve decreases the sensitivity of our spectra towards shorter wavelengths, which can introduce a color term in the binned spectra.  An analysis of bright objects in our sample indicates that the flux in the shortest wavelength bin, from 3200--3400 \AA, is typically overestimated from binning and is reduced by 4\% in each spectrum.  The relative flux calibration in other bins is accurate to better than 3\% and the absolute flux calibration to 5\%, when compared with the average flux in the bin calculated from spectra extracted at the full resolution of the instrument.

Figure 1 demonstrates this technique on 2MASS J12073347-3932540.  The highlighted regions are dominated by continuum emission; other regions are dominated line emission, especially the Balmer series and \ion{Ca}{2} H \& K.  We binned the image over these highlighted wavelength intervals to obtain accurate measurements of the continuum while minimizing the contamination by strong lines.  The squares show the resulting spectrum.  

For spectral templates, we use a spectrum of the M7 field dwarf LHS 3003 observed here, a published spectrum of the non-accreting M5.5 dwarf MHO 7 in Taurus \citep{Her08}, and a Palomar Double Spectrograph spectrum of the young M4 star LkCa 1 in Taurus (in preparation).  A scaled photospheric template is subtracted from the spectrum before measuring emission in  \ion{He}{1}, [\ion{O}{1}], and \ion{Ca}{2} IR triplet lines.  Emission in the Balmer and \ion{Ca}{2} H \& K lines is measured directly from the spectrum.  The spectral templates are also used to estimate photospheric emission from stars with excess continuum emission.

Emission lines were fit with a Gaussian profile to measure line equivalent widths and fluxes.  For the six faint targets USco J160603.75-221930.0, SSSPM J1102-3431, DENIS-P J160603.9-205644, UScoCTIO 112, Oph J162225-240515A, and Oph J162225-240515B, the equivalent width of lines at $<5500$ \AA\ are calculated by using the very low-resolution continuum measurements described above to estimate the correct continuum level at the relevant wavelength.

\section{Analysis}

\subsection{Spectral and Physical Properties of Sample}
Fig.~2 shows the 5600--9000 \AA\ spectra of each of our targets.  Optical spectral types were estimated by direct comparison of the red optical
spectra of our targets to the young M dwarf spectral
standards provided by K.Luhman.  We adopted the spectral type of the standard that provided the best match to the entire optical spectral shape of our targets, which is determined primarily by the depth of VO and TiO absorption features.  This method results in uncertainties of $\sim 0.25-0.5$ subclasses.

The relative gravity of M-dwarfs can be characterized by the absorption depth in the \ion{Na}{1} $\lambda8183,8195$ and \ion{K}{1} $\lambda7665,7699$ lines.  \citet{Sle06} developed a gravity-sensitive index comparing the \ion{Na}{1} doublet absorption depth to the nearby continuum.  Shallow absorption is seen in these features, with \ion{Na}{1}-8189 indexes that range from 0.87--0.95, which is consistent with the intermediate gravities typical for M-dwarfs in USco and the TWA.  The older M7 template, LHS 3003, has much deeper absorption, with a \ion{Na}{1}-8189 index of $0.75$ that is typical of main-sequence dwarfs.

2MASS J12073347-3932540 and SSSPM J1102-3431 are confirmed members of the nearby  TWA, which at an age of $\sim 10$ Myr is no longer associated with molecular material \citep{Giz02,Sch05,Mam05}.  Relative to spectral standards, both of our TWA targets have $A_V\sim0.0$ based on the optical extinction law from \citet{Car89} and a selective-to-total extinction parameter $R_V=3.1$.  The other brown dwarfs belong to the 5 Myr old USco, which suffers from moderate extinction of $A_V<2$ \citep{deb99,Pre02}.  The extinctions to USco J160723.82-221102.0 and USco 160603.75-221930.0 are $A_V\sim0.3\pm0.3$ and $A_V\sim0.0\pm0.3$, respectively. 

Table 3 lists the physical properties of our sample, derived from the effective temperature based on the optical spectral type \citep{Luh03}, extinction, J-band photometry \citep{Web99,Ard00,Giz02,Mar04,Sch05,Luh07,Lod08}, the near-IR extinction law from \citet{Rie85}, and pre-main sequence evolutionary tracks from \citet{Bar98} and \citet{Cha00}.  Parallax distances are used for TW Hya, 2MASS J12073347-3932540, and SSSPM J1102-3431 \citep{Wic98,Duc08,Tei08}.  A kinematic distance of 34 pc is used for Hen 3-600 \citep{Mam05}.  For USco stars we assume the cluster distance of 145 pc \citep{deZ99}.  The luminosity of Upper Sco brown dwarfs are generally lower than predicted for 5 Myr old objects from pre-main sequence evolutionary tracks but are consistent with the luminosities of other very low-mass members of Upper Sco \citep[see discussion in][]{Luh07}.

\subsection{Identifying Accretion}
The spectrum from 3200--9000 \AA\ of a young brown dwarf consists of a red photospheric continuum and weak chromospheric emission in the Balmer and \ion{Ca}{2} H \& K lines.  The presence of accretion alters the spectrum by producing excess line and continuum emission, and is typically identified by the strength or shape of line emission (some prominent lines are listed in Table 2) or by the veiling of the photospheric continuum by the accretion continuum \citep[e.g.,][]{Whi03,Muz03,Nat04,Moh05}.  The accretion continuum studied here consists of the bright Balmer continuum at $<3646$ \AA, the fainter Paschen continuum emission at $<8200$ \AA, and an H$^-$ continuum that spans our entire wavelength region.  
The excess continuum emission is easiest to detect at $<3646$ \AA, where the contrast between the bright Balmer plus Paschen continuum and the faint photospheric continuum from the brown dwarf is high.  The fainter Paschen continuum emission is easiest to detect at 4000 \AA, where the photospheric continuum is weaker relative to that at longer wavelengths and the flux is not contaminated by line emission.  

The Balmer jump at 3646 \AA\ (shifted to 3700 \AA\ in low-resolution spectra because high Balmer lines blend together) is characteristic of accretors (Kuhi 1966; Valenti et al.~1993; Gullbring et al.~1998; Calvet \& Gullbring 1998; Herczeg \& Hillenbrand 2008).  The size of the observed Balmer jump ($BJ_{obs}$ in Table 3) is defined here as the ratio of observed emission (photospheric plus accretion) at 3600 \AA\ divided by that at 4000 \AA.  For the faint stars in our sample, the ratio is calculated using the binned data at the points closest in wavelength to the nominal values.

Figure 3 shows the blue spectra of three low-mass stars and the eight of our nine brown dwarfs in our sample which have blue spectra.  The spectra of the brighter targets reveal that the Balmer jump is detected from 2MASS J12073347-3932540, USco CTIO 128, TW Hya, and Hen 3-600A but is not detected from Hen 3-600B.  The spectra of the six fainter targets is too noisy at $R\sim700$ to determine whether the Balmer jump is present.  Figure 4 shows the binned spectra of those six faint targets compared with a photospheric template.  For UScoCTIO 112 we use the young M5.5 dwarf MHO 7 as a template.  For the M7-M8.75 brown dwarfs we use the active M7 star LHS 3003 as a template.  We roughly corrected for the temperature difference between M7 and M9 objects by multiplying the observed spectrum by relative strength of blackbody spectra of 2650 K for an M7 dwarf \citep{Luh03} and the photospheric temperature of each object.  The blue continuum from SSSPM J1102-3431, DENIS-P J160603.9-205644, UScoCTIO 112, Oph J162225-240515B, and Oph J162225-240515A falls rapidly toward shorter wavelenths, similar to the photospheric template.  Any veiling due to accretion is undetected at $>4000$ \AA\ in these objects.  In contrast, USco J160603.75-221930.0 has a relatively flat spectrum from 4000--5000 \AA, consistent with  high veiling.

SSSPM J1102-3431, Oph J162225-240515B, DENIS-P J160603.9-205644, and USco J160603.75-221930.0 have excess emission at $<3700$ \AA, which is attributed to the Balmer continuum.  The observed Balmer jump for these targets are larger than the observed Balmer jump of 0.35--0.45 for chromospherically-active M-dwarfs \citep{Her08}.  We classify these four stars as accretors based on excess Balmer continuum emission.  The Balmer jumps of $\sim 0.4$ and $<0.4$ for UScoCTIO 112 and USco J160723.82-221102.0, respectively, are lower than expected from a late-M brown dwarf photosphere and suggest that they are not accreting.

\subsection{Measuring Accretion}
The excess accretion continuum emission described above is produced near where the accreting gas shocks at the brown dwarf surface \citep{Har91,Val93,Cal98}.  With a bolometric correction, the observed excess flux over a narrow wavelength range can be converted into a total accretion continuum luminosity, $L_{acc}$, which is roughly the amount of energy released by gas that accretes onto the star.  We can then calculate the accretion rate, $\dot{M}$, from $L_{acc}$  by
\begin{equation}
\dot{M}=\frac{1.25 R_* L_{acc}}{G M_*},
\end{equation}
for a star of radius $R_*$, mass $M_*$, and disk truncation radius of $5$ $R_*$ \citep{Gul98}.  The shape of the accretion continuum is calculated by modelling a pure hydrogen, isothermal, plane-parallel slab \citep{Val93,Her08}.   The excess Balmer continuum flux from 3200--3600 \AA\ is then converted to a total accretion luminosity by scaling the model spectrum to the measured excess emission.   These models are simplistic and less physical than the shock models developed by \citet{Cal98} but provide similar bolometric corrections and consistency with most previous similar accretion rate measurements.

The bolometric correction depends on the size of the Balmer jump of the accretion continuum.  This Balmer jump is larger than the observed Balmer jump, which is reduced in size by the photospheric emission.  The observed Balmer jump is determined by the combination of photospheric emission, which has a ratio of flux at 3600 \AA\ to 4000 \AA\ of 0.35--0.45, and by the accretion continuum, which has that flux ratio typically $>2$.  The Balmer jump of the accretion continuum can be tuned in the models by changing the temperature and density of the slab.  We assume a slab temperature of $9000$ K and a length for the slab of $2\times10^7$ cm$^3$.  The density of the slab then determines the opacity and thereby the Balmer jump.
The Balmer jump of the accretion continuum ($BJ_{acc}$ in Table 3) can be constrained for most accretors and is used to determine the bolometric correction. 
Uncertainty in the size of the Balmer jump of the accretion continuum introduces a $\sim 25$\% uncertainty in accretion luminosity.
\footnotetext[3]{These values are approximately consistent with those calculated by \citep{Her08} and \citet{Val93}, with a temperature slightly lower than the $10^4$ K used by \citet{Gul98}.  These assumptions introduce small differences in the bolometric correction and therefore the accretion luminosity.}

Table 3 lists the accretion luminosity and rate (or upper limit) for each source.
The luminosity in Balmer lines is equivalent to that in the accretion continuum but is excluded in measurements of the total accretion luminosity for consistency with previous accretion rate estimates.   A total uncertainty of 3--5 dex is adopted for accretion rate measurements calculated in this paper, which incorporates uncertainties in distance and extinction and systematic uncertainties in pre-main sequence tracks, bolometric corrections, the exclusion of emission lines.  \citet{Her08} discusses more thoroughly how these uncertainties affect the mass accretion rate.

\subsection{COMMENTS ON INDIVIDUAL SOURCES}

\subsubsection{2MASS J12073347-3932540}
\citet{Giz02} identified 2MASS J12073347-3932540 as a brown dwarf member of the TWA.  The strength and shape of Balmer and \ion{He}{1} lines indicate that accretion is ongoing \citep[e.g.][]{Giz02,Moh03}.  We use our spectra to adopt a spectral type of M8.25.  The excess Balmer continuum emission is consistent with the classification as an accretor and leads to an accretion rate of $1.3\times10^{-12}$ \mdotyr.  We also detect strong emission in the Balmer lines, \ion{Ca}{2} H \& K, and several \ion{He}{1} lines.  The \ion{Ca}{2} IR triplet is marginally detected, with an equivalent width of $0.5$ \AA\ in each of the three lines.

 The accretion luminosity measured here is similar to that measured in Nov. 2006 and Feb. 2007 by \citet{Her08} using the same method.  For comparison, the accretion rates from \citet{Her08} should be multiplied by 0.6 to account for a change in the adopted stellar parameters. The flux in higher Balmer lines and \ion{He}{1} $\lambda6678$ are also similar in all three observations.  However, the H$\alpha$ line flux is three times smaller than that observed in Feb. 2007.  The \ion{He}{1} $\lambda5876$ line luminosity changes by a factor of 2 over the three observations and is correlated with the $\sim 30\%$ difference in accretion rate.  We also detect [\ion{O}{1}] emission, an outflow diagnostic, with a flux larger than that detected in two previous observations \citep{Her08}.  We infer that the changes in the strength in H$\alpha$ and [\ion{O}{1}] emission do not always correlate with changes in accretion rate. 

That the UV-excess measure of accretion onto 2MASS J12073347-3932540 is similar in three epochs could indicate that the accretion may be relatively steady.  However, the V-band magnitude of 2MASS J12073347-3932540 reported in the literature varies between 19.5--20.5, despite that the I-band magnitude is relatively stable \citep{Giz02,Duc08,Koe08}.  Such variability is likely directly attributable to accretion variability and implies that the accretion rate can range from $10^{-12}-10^{-11}$ $M_\odot$ yr$^{-1}$.  Changes in emission line fluxes and profile shapes from 2MASS J12073347-3932540 have been commonly detected, with at least some of the differences attributable to a varying accretion rate \citep{Sch06,Ste07}.
With only three observations, we may not be sensitive to such accretion variability.  This explanation and the magnitude of accretion rate changes should be tested with photometric or spectroscopic monitoring of U-band emission.

\subsubsection{SSSPM J1102-3431}
\citet{Sch05} identified SSSPM J1102-3431 as a brown-dwarf member of the TWA.  We adopt a spectral type of M8.5.  
The excess Balmer continuum emission detected here leads to an accretion rate of $1.6\times10^{-13}$ \mdotyr, the lowest accretion rate measured to date.  The H$\alpha$  equivalent width of 50 \AA\ measured here and our detection of \ion{He}{1} line emission also indicate ongoing accretion, as does the 64 \AA\ equivalent width measured by \citet{Loo07}.  This brown dwarf had not been previously classified as an accretor because the H$\alpha$ 10\% width of 194 \kms\ \citep{Ste07} is intermediate between accretors and non-accretors for late M-dwarfs \citep{Whi03,Muz05,Moh05}.  Ongoing accretion is consistent with the presence of a disk \citep{Mor08,Ria08}.  Emission in the [O I] $\lambda6300$ line indicates the presence of an outflow.

\subsubsection{UScoCTIO 112}
\citet{Ard00} identified UScoCTIO 112 as an M5.5 in Upper Sco.  According to evolutionary tracks mass of UScoCTIO 112 is slightly larger than the stellar/subtellar boundary, though we group it with the brown dwarfs for ease of terminology.  The emission at $<3700$ \AA\ is consistent with the photospheric emission from a non-accreting young M5.5 dwarf.  An upper limit on the veiling of $\sim 0.4$ leads to an upper limit on accretion rate of $6\times10^{-13}$ \mdotyr.  The H$\alpha$ equivalent with of $22$ \AA\ is intermediate between accretors and non-accretors for late M-dwarfs and cannot be used to infer the presence or absence of accretion.  However, our non-detection of accretion onto UScoCTIO 112 is consistent with the narrow H$\alpha$ 10\% width of $\sim 110$ \kms\ \citep{Muz03,Moh05}.  The lack of accretion is also consistent with non-detections of emission in the \ion{He}{1} and \ion{Ca}{2} IR triplet lines.  However, we find some weak emission in the [\ion{O}{1}] $\lambda6300$ line, which could indicate the presence of an outflow.

A disk around UScoCTIO 112 has been identified from the presence of an L-band and mid-IR excesses \citep{Jay03,Sch07}.  Our stringent upper limit on the accretion rate indicates that despite the presence of a disk, little or no material is accreting onto the star.

\subsubsection{UScoCTIO 128}
\citet{Ard00} identified UScoCTIO 128 as an M7 brown dwarf member of Upper Sco.   The excess Balmer continuum emission leads to an accretion rate of $4\times10^{-12}$ $M_\odot$ yr$^{-1}$.  UScoCTIO 128 had been identified as an accretor by \citet{Ard00} based on an H$\alpha$ equivalent width of 130.5 \AA.  However, \citet{Muz03} classified UScoCTIO 128 as a non-accretor based on the H$\alpha$ 10\% width of 193 \AA, though noted that emission in \ion{He}{1} and \ion{Ca}{2} triplet lines may indicate some accretion.  Their measurement of an H$\alpha$ equivalent width of 60 \AA\ is sufficient to identify UScoCTIO 128 as an accretor.  \citet{Moh05} detected an IR excess but also classified UScoCTIO 128 as a non-accretor based on an H$\alpha$ equivalent width of 15.9 \AA\ and 10\% width of 121 \kms.  The \ion{He}{1} $\lambda6678$ line was detected, but with an equivalent width of 0.3 \AA. 

Our detections of strong emission in the Balmer lines, including an H$\alpha$ equivalent width of 105 \AA, \ion{He}{1} lines, and the \ion{Ca}{2} IR triplet (equivalent widths of $1.2$ \AA, or fluxes of $2.8\times10^{-16}$ erg cm$^{-2}$ s$^{-1}$) all indicate accretion at a level similar to that when the target was observed by \citet{Ard00} and \citet{Muz03}.  On the other hand, \citet{Moh05} must have observed UScoCTIO 128 during a period of weak accretion, though their detection of \ion{He}{1} line emission suggests that some accretion was ongoing.

\subsubsection{USco J160603.75-221930.0}
\citet{Lod08} identified USco J160603.75-221930.0 as a brown-dwarf member of Upper Sco.  This brown dwarf was selected for our study because it has a near-IR spectral type of L2, the latest BD in the \citet{Lod08} sample.  We find an optical spectral type of M8.75.  The excess Balmer continuum emission leads to an accretion rate of $5\times 10^{-13}$ \mdotyr.  The H$\alpha$ equivalent width of $\sim750\pm80$ \AA\ is one of the largest detected for a young star and is at least partially explained by the high contrast against the weak photopsheric emission of an M8.75 brown dwarf.  No \ion{Ca}{2} IR triplet or [\ion{O}{1}] emission is detected from USco J160603.75-221930.0. 


\subsubsection{USco J160723.82-221102.0}
USco J160723.82-221102.0 was identified by \citet{Lod08} as a brown-dwarf member of Upper Sco and was selected for our study because it was the brightest target that was classified as a young L1 in their sample.  The optical spectral type of USco J160723.82-221102.0 is M8.5, about 3 spectral types earlier than the near-IR spectral type.  Our observation of USco 160723.82-221102.0 is not sensitive to either photospheric emission at $<5500$ \AA\ or excess Balmer continuum emission.  The H$\alpha$ equivalent width of 23 \AA\ is indeterminant for assessing the presence or absence of accretion.  The equivalent width is comparable to the most chromospherically active of M8 stars \citep{Moh03}, which is consistent with youth.  No other emission lines are detected.   We therefore suggest that USco J160723.82-221102.0 is not accreting.  Based on the non-detections of the \ion{He}{1} $\lambda5876$ and $\lambda6678$ lines$^4$, the upper limit on accretion rate is $5\times10^{-12}$ \mdotyr.
\footnotetext[4]{Using the formulae $\log L_{acc}=6.7 + 1.39 \log L_{6678}$ and  $\log L_{acc}=5.3 + 1.26 \log L_{5876}$, where $L_{acc}$ and $L_{line}$ are in units of $L_\odot$.  These two relationships are calculated from line and accretion luminosities presented here and in \citet{Her08}.}

\subsubsection{Oph J162225-240515A}
Oph J162225-240515A was identified as the primary component of a $1\farcs9$ binary in USco by \citet{All05}.  \citet{Luh07} classified the primary as M7$\pm0.25$, which is adopted here.   As with USco J160723.82-221102.0, the H$\alpha$ equivalent width from Oph J162225-240515A of 19 \kms\ \citep{Jay06} is indeterminant between accretion and non-accretion.  The {\it Spitzer} IRAC colors obtained by \citet{Clo07} are consistent with a slight IR excess above photospheric emission when compared to relative to photospheric colors of young brown dwarfs in $\sigma$ Ori \citep{Luh08}, though the excess may not be significant because extracting IRAC photometry of close binaries is difficult.

Some photospheric emission is detected at 4000 \AA\ but undetected at $<3700$ \AA.  This non-detection leads to an upper limit on the observed Balmer jump of 0.4 \AA, which is consistent with the absence of accretion.  The upper limit on accretion rate of $\dot{M}<5\times10^{-14}$ $M_\odot$ yr$^{-1}$ is calculated from the upper flux limit at 3600 \AA\ by assuming a veiling at 3600 \AA\ of $<1$.

\subsubsection{Oph J162225-240515B}
Oph J162225-240515B is the secondary component of Oph J162225-240515A.   \citet{Luh07} classified the primary as M8.75$\pm0.25$, which is adopted here.  The H$\alpha$ equivalent width of 65 \AA\ \citep{Jay06} and the excess emission in the {\it Spitzer} IRAC imaging \citep{Clo07} indicate that this component is accreting from a circumstellar disk.

The excess Balmer continuum emission is calculated from the flux measurement at 3600 \AA, which is significant to $3.7\sigma$.  The accretion rate of $\sim 2.5\times10^{-12}$ $M_\odot$ yr$^{-1}$ is uncertain because the Paschen continuum emission is weaker than our detection limit, so the size of the Balmer jump is uncertain.   Oph J162225-240515B also has an H$\alpha$ equivalent width of 260 \AA\ and several \ion{He}{1} lines seen in emission, which are consistent with accretion.  The \ion{Ca}{2} $\lambda8542$ and $\lambda8662$ lines have tentative detections with fluxes of  $7\times10^{-17}$ erg cm$^{-2}$ s$^{-1}$, or equivalent widths of $\sim 2$ \AA.

\subsubsection{DENIS-P J160603.9-205644}
DENIS-P J160603.9-205644 was identified as a brown dwarf member of USco by \citet{Mar04}.  We classify DENIS-P J160603.9-205644 as an M7.25 brown dwarf.  \citet{Mar04} and \citet{Moh05} identified accretion onto DENIS-P J160603.9-205644 by the large H$\alpha$ equivalent width of 70--105 \AA\ and a 10\% width of 306 \kms.  From the excess UV emission we measure an accretion rate of $2.5\times10^{-12}$ \mdotyr.  The spectrum also shows emission in the higher H Balmer lines, several \ion{He}{1} lines.  All three \ion{Ca}{2} IR triplet lines are detected with fluxes $\sim 7.5\times10^{-16}$ erg cm$^{-2}$ s$^{-1}$, or equivalent widths of 2.0 \AA.  No emission is detected in the [\ion{O}{1}] $\lambda6300$ line.

\subsubsection{Hen 3-600A}
Hen 3-600A is the M3 primary component of the $1\farcs44$ binary of Hen 3-600, which was identified as a member of the TWA by \citet{del89}.   Both accretion and a disk have been detected from Hen 3-600A \citep{Muz00,Jay99,Met04}.  The excess Balmer continuum emission indicates accretion at a rate of $2.5\times10^{-10}$ \mdotyr, larger than the rate of $5\times10^{-11}$ as determined from modeling the H$\alpha$ line profile \citep{Muz00}.  The [\ion{O}{1}] emission indicates the presence of an outflow.  The upper flux limit on the \ion{He}{1} $\lambda6678$ line is consistent with the small accretion luminosity.


\subsubsection{Hen 3-600B}
Hen 3-600B is the M3.5 secondary component of Hen 3-600.  Neither accretion nor a disk have been previously detected from Hen 3-600B.  We do not detect any excess Balmer continuum emission from Hen 3-600B.  An upper limit of 0.3 on the veiling at 3600 \AA\ leads to an upper limit on accretion of $<8\times10^{-12}$ \mdotyr.  Hen 3-600B exhibits weak H Balmer and \ion{Ca}{2} H \& K line emission, but no other emission lines are detected.


\subsubsection{TW Hya}
TW Hya was first identified as a K7 classical T Tauri star by \citet{Ruc83} based on excess Balmer continuum emission, strong line emission, and excess IR emission.  \citet{del89} found that TW Hya is associated with other nearby stars, which became the TWA.  We measure an accretion rate of $8\times10^{-10}$ \mdotyr, which is about a factor of two smaller than previous UV-excess measures of accretion \citep{Her04,Her08}, and closer to that measured by \citet{Muz00} from H$\alpha$ line modelling.   In contrast, the Balmer line emission detected here is stronger than during the two larger accretion rate measurements.

\section{DISCUSSION}
\subsection{Sensitivity to Tiny Accretion Rates}

Accretion signatures are detected from 2MASS J12073347-3932540, SSSPM J1102-3431, USco J160603.75-221930.0, Oph J162225-240515B, UScoCTIO 128, DENIS-P J160603.9-205644, TW Hya, and Hen 3-600A and and absent from USco J160723.82-221102.0, Oph J162225-240515A, UScoCTIO 112, and Hen 3-600B.  With the exception of USco J160723.82-221102.0, the accretion rates and upper limits are measured from the excess Balmer continuum emission shortward of $3600$ \AA.  For UScoCTIO 112, SSSPM J1102-3431, DENIS-P J160603.9-205644, USco J160603.75-221930.0, Oph J162225-240515B, and Oph J162225-240515A, the blue and U-band continuum emission is faint and measured only after binning the image in the dispersion direction and subsequently measuring the counts.  Extracting very low-resolution continuum spectra increases our sensitivity to low continuum fluxes by reducing the uncertainy in background subtraction.

The sensitivity of UV-excess measurements to small accretion luminosities relies on the ability to detect the Balmer continuum in excess of the photospheric emission.  At earlier spectral types, this limit is determined by the strength of the photospheric emission and an upper limit on the veiling, typically $\sim 0.1-0.3$.  The upper limits on accretion rate for the M3.5 star Hen 3-600B and the M5.5 brown dwarf UScoCTIO 112 are $\sim 8\times 10^{-12}$ $M_\odot$ yr$^{-1}$ and $6\times10^{-13}$ $M_\odot$ yr$^{-1}$, respectively.   Since these limits depend on the contrast ratio to the photospheric continuum, which is proportional to $R^2$, and since $\dot{M}\propto R$, for two non-accreting stars with the same mass, one with twice the radius will have an upper limit on accretion rate that is eight times larger.

Towards later spectral types, the fainter and redder photospheric emission becomes undetectable at short wavelengths.  The telescopic sensitivity then sets the limit on a detectable accretion luminosity.  Our very low-resolution continuum spectra from the binned data are sensitive to $26.3$ mag at 3600 \AA\ in a 900s integration.  This sensitivity limit corresponds to an accretion rate of $\sim 10^{-13}$ $M_\odot$ yr$^{-1}$ for a 0.02 $M_\odot$, $0.2$ $R_\odot$ brown dwarf at 140 pc with $A_V=0$.

The faint excess Balmer continuum emission detected here demonstrates that UBVRI spectrophotometry can also be used to measure small accretion rates onto brown dwarfs.  \citet{Gul98} derived a relationship between dereddened excess U-band emission and accretion luminosity, which is especially useful to measure accretion onto targets in crowded  fields or faint targets \citep[e.g.][]{Whi01,Rob04,Rom04}.  When this relationship is applied to brown dwarfs, the excess U-band emission needs to be corrected for strong Balmer line emission.  For 2MASS J12073347-3932540, USco J160603.75-221930.0, and SSSPM J1102-3431 the Balmer lines contribute $\sim 40-60$\% of the U-band photons.  
 The photospheric colors should be obtained from VIJHK bands, since continuum emission in the B and R bands may be significantly contaminated by emission in Balmer lines.  No significant adjustment is needed for low-mass stars because the Balmer and \ion{Ca}{2} H \& K line luminosities are weak relative to the total accretion continuum luminosity, contributing only 10\% of the U-band photons from TW Hya.  The stronger emission in the Balmer lines relative to the continuum luminosity is consistent with the lower optical depth in the accretion flow that \citet{Gat06} inferred for small accretion rates, based on Pa$\beta$ to Br$\gamma$ ratios.

In the absence of spectra below $3700$ \AA, emission line luminosities can be used to measure accretion rates \citep{Muz98,Nat04,Moh05,Her08}.  The most prominent  diagnostic of accretion is H$\alpha$ emission.  Figure 5 shows that the H$\alpha$ equivalent width is correlated with accretion rate, though with large scatter in the $L_{acc}-L_{H\alpha}$ relationship is large.  Very large equivalent widths are commonly produced by accretion onto cooler objects because the photospheric continuum is both weaker and redder compared with hotter objects.  The width of the H$\alpha$ line is also a useful indicator for accretion, although it can be indeterminant for intermediate values, such as those measured from both UScoCTIO 128 and SSSPM J1102-3431.  For these two brown dwarfs, the H$\alpha$ equivalent widths of $\sim50-100$ \AA\ are sufficient to categorize them as accretors.

Instead of using H$\alpha$, the luminosity in higher Balmer lines and \ion{He}{1} lines provide more accurate estimates of accretion luminosity \citep{Her08}.  For Oph J162225-240515A, the detected H$\gamma$ flux, which is likely chromospheric, yields $\log L_{acc}<-6.6$, which is comparable to our upper limit from the Balmer continuum.  Since we lack a blue spectrum for USco J160723.82-221102.0, we rely on \ion{He}{1} $\lambda5876$ and $\lambda6678$ to arrive at $\log L_{acc}<-5.1$, which is substantially less sensitive than upper limits that could be calculated by measuring the flux in the Balmer continuum or high Balmer lines.  The upper limit on accretion rate obtained from the \ion{Ca}{2} IR triplet is even higher, at $\log L_{acc}<-4$, because \ion{He}{1} is detected from a higher percentage of low-mass accretors than is \ion{Ca}{2} emission.

The observations described in this paper provide an instantaneous snapshot of the accretion properties of our sample but do not necessarily represent the time-averaged accretion rates.  Variability is a fundamental aspect of accretion, with changes in continuum flux or line fluxes seen on timescales of hours to years \citep[e.g.][]{Her94,Joh95,Ale02,Sch06,Ste07,Gro07,Koe08}.   However, as seen in our three observations of 2MASS J12073347-3932540, optically-thick line fluxes may sometimes change without corresponding changes in accretion rate.  In monitoring of several CTTSs at high spectral resolution, \citet{Gah08} found significant changes in line fluxes without corresponding changes in accretion rate.  Long-term photometric monitoring of young stars is required to understand accretion variability across the range of masses and mass accretion rates.


\subsection{Optical spectral types of two USco objects}

The optical spectral types of the USco J160723.82-221102.0 (M8.5) and USco J160603.75-221930.0 (M8.75) are several sub-classes earlier than the near-IR classification of L1 and L2, respectively \citep{Lod08}. 
That two brown dwarfs from the \citet{Lod08} sample, including the latest L-dwarf, have optical spectral types consistent with late M-dwarfs may indicate that the other 15 L0-L1 dwarfs in their sample also have optical spectral types of late-M.  The mass of an object is usually calculated by comparing the temperature, here obtained from the temperature-spectral type relation of \citet{Luh03}, and luminosity to pre-main sequence evolutionary tracks \citep{Bar98,Cha00}. Relative to an L1 spectral type, the hotter temperature inferred from an M8.75 spectral type would increase the lower mass range sampled from the IMF in Upper Sco to $\sim 0.02-0.03$ $M_\odot$ from $0.008$ $M_\odot$.  On the other hand,  several candidate L-dwarfs in \citet{Lod08} are fainter and redder in JHK than the two targets selected here and may indeed be early L-dwarfs.

In a study of the 2--3 Myr old cluster IC348, \citet{Luh03} similarly found that spectral classifications obtained from low-resolution near-IR spectra were systematically 1--2 subclasses later than those measured from optical spectra, and attributed some of the discrepancy to different strengths of steam band absorption for low-gravity young objects and high-gravity dwarfs.  The \citet{Luh03} relationship between temperature and spectral type applies to optical spectral types, but no similar relation has been derived for near-IR spectral types.  Near-IR spectra are not well-correlated with photospheric temperature for field late M-L dwarfs because the shape of near-IR spectra are more sensitive than optical spectra to differences in gravity, dust opacities, metallicities, and binarity \citep{Kna04,Bur08}.   We caution against the use of near-IR spectral types to obtain temperatures and masses of young brown dwarfs until these effects are better characterized.

\section{Acknowledgements}
We thank Adam Kraus for valuable discussion of the proposal and our results.  We are grateful to Kevin Luhman for providing spectra for his spectral sequence of young late-M brown dwarfs and for classifying the TWA BDs 2MASS J12073347-3932540 and SSSPM J1102-3431 from our data.  GJH thanks Leonardo Testi for discussion of the optical depth in hydrogen lines.  We also thank Kevin Luhman, Nicolas Lodieu, and the anonymous referee for helpful comments.

The data presented herein were obtained at the W.M. Keck Observatory, which is operated as a scientific partnership among the California Institute of Technology, the University of California and the National Aeronautics and Space Administration. The Observatory was made possible by the generous financial support of the W.M. Keck Foundation.  K. L. C. is supported by NASA through the Spitzer Space Telescope Fellowship Program, through a contract issued by the Jet Propulsion Laboratory, California Institute of Technology under a contract with National Aeronautics and Space Administration.

\section{Appendix A: Background Visual Companions}
Our observations with the guider revealed close visual companions to USco J160723.82-221102.0 and USco J160603.75-221930.0.  In both cases the intended brown dwarf target was brighter in the near-IR but fainted in the optical.  USco J160723.82-221102.0B is located at $2\farcs2$ from USco J160723.82-221102.0 at a PA of 90$^\circ$.  USco J160723.82-221102.0B is an M3 star with an H$\alpha$ equivalent width of $<1.5$ \AA, which is consistent with a field star.  The extinction is $A_V\sim0.5$ to USco J160723.82-221102.0B.  For USco J160723.82-221102.0B to be associated with USco J160723.82-221102.0 would require gray extinction of $A_V\sim8.5$.  USco J160723.82-221102.0B is likely a background star.  
USco J160603.75-221930.0B is located $3\farcs3$ from the brown dwarf USco J160603.75-221930.0A at a PA of $\sim105^\circ$.  USco J160603.75-221930.0B is a background A-star.

2MASS J1207B has a planetary-mass companion located $0\farcs7$ from the primary.  The companion is much fainter than the primary and not resolvable.  Hen 3-600 and Oph 162225-240515 are both binary systems with components that are physically associated with each other and are discussed in detail in our paper.

\begin{table}[!h]
\caption{Observation Log}
\label{tab:obsx.tab}
{\footnotesize \begin{tabular}{cccccccccc}
\hline
Star        &  $t_{exp}^a$ (s)\\ 
\hline
\hline
USco J160723.82−221102.0   & 1200   \\
USco J160603.75-221930.0 & 1800 \\
SSSPM J1102-3431  & 2700     \\
2MASS J12073347-3932540            & 900      \\
Oph J162225-240515     & 900     \\
DENIS-P J160603.9-205644   & 600 \\
UScoCTIO 128 & 1200 \\
TW Hya      &   20   \\
Hen 3-600  &  210   \\
LHS 3003$^b$  & 360\\
\hline
\multicolumn{2}{l}{$^a$Exposure time for blue channel.}\\
\multicolumn{2}{l}{$^b$Used as a spectral template.}\\
\end{tabular}}
\end{table}





\begin{table}[!h]
\caption{List of Selected Prominent Lines$^a$}
{\scriptsize \begin{tabular}{cccccccccc}
\hline
Object & H$\alpha$ & H$\beta$ & H$\gamma$ & H$\delta$ & H-6 & \ion{Ca}{2} K & \ion{He}{1} $\lambda5876$ & \ion{He}{1} $\lambda6678$ & [\ion{O}{1}] $\lambda6300$ \\
\hline
\hline
TW Hya         &   280    & 64      & 37        & 30    & 30     &  19   & 3.1$^b$ &  1.05    & $<0.9$\\
               & -10.26    & -11.33  & -11.74   & -11.93 & -12.08 & -12.27 & -12.38$^b$ & -12.80 & $<-12.8$\\
\hline
Hen 3-600A    & 37    &  20    &  25    &  23    &  25      & 16     &  0.77  & $<0.6$ & 0.54 \\
              & -11.67 & -12.32 & -12.45 & -12.54 & -12.70  &-12.91 & -13.20 & $<-13.7$  & -13.77\\
\hline
Hen 3-600B   & 3.4    &  4.9  &    3.4    &  3.5   & 4.5   &  18.2  & $<1.3$  & $<0.7$  & $<0.3$\\
             & -12.91 & -13.16 &  -13.49  & -13.61 & -13.8 & -13.19 & $<-13.7$ & $<-13.7$ & $<-14.2$\\
\hline
UScoCTIO 112 & 21     & 16      &  17     & 14     & 14     & 25    & $<1.1$  & $<0.2$   & $0.5$\\
             & -14.41 &  -15.22 & -15.53  & -15.73 & -15.98 &-15.71 & $<-16.2$ & $<16.6$ & 16.34\\
\hline
UScoCTIO 128 & 102    & 160     & 230   & 200    & 120    & 54  &     24 &  6.0    &  $<1.3$\\
             & -14.27  & -14.64 &-14.76 & -14.87 & -15.08 & -15.52 & -14.37 & -15.72  &  $<-16.52$\\
\hline
2MASS J12073347-3932540  & 126   &  194    & 260    & 230  & 130 & 32   &  29    &   5.9      & 3.7\\
               & -14.03    & -14.52  & -14.68 & -14.79     & -14.97 &  -15.64 & -15.26 & -15.61 & -15.99\\
\hline
SSSPM J1102-3431 & 50        &  95     &  150    &   130  &  120   &   100   & 7.5  & 1.7         & 4.3\\
                 & -14.62    &  -15.27 & -15.53  & -15.70 & -15.86 & -15.91 & -16.12 & -16.32 & -16.13\\
\hline
USco J160603.75-221930.0 &   760    & 330     & 270   &  170  &   83   & 25    & 46       & 8.2      & $<15$\\ 
                         &  -14.56   &  -15.48 &-15.75 & -15.89& -16.10 & -16.61 & -16.32 & -16.76 & $<-16.6$\\
\hline
USco J160723.82−221102.0&   11.5    & -- & -- & --   & --& --  &   $<0.6$  & $<0.4$ & $<20$\\
                        &   16.15   & -- & -- & --   & -- &--  &   $<-17.0$& $<16.9$ & $<-16.4$\\
\hline
Oph J162225-240515A &  20  &  11     &  38    &   19      &    3    &  40      &  $<3.8$    & $<1.3$ & $<2.4$\\
               &  -15.49   & -16.26  & -16.62 & -16.92$^c$ & -18.1$^c$ & -16.80 & $<-16.7$ & $<-16.7$ & $<-16.6$ \\  
\hline
Oph J162225-240515B&  270  &  360    &  410   &  320   &    160  &  190    & $<40$ &  $<5$  & $<20$ \\
               &  -15.02   & -15.66  & -16.08 & -16.24 &  -16.53 &  -16.46 & $<-16.3$  & $<-17$ & $<-16.6$\\
\hline
DENIS-P J160603.9-205644&   131      &  93     & 110     & 70     & 63     & 170 &  4.5  &   2.4      &  $<2$\\
                         &  -14.08   &  -15.01 &  -15.43 & -15.64 & -15.80 & -15.36 & -16.07  & -16.06  & $<-16.3$\\
\hline
\hline
\multicolumn{7}{l}{$^a$Top line for each object is the line equivalent width in \AA}\\
\multicolumn{7}{l}{~~~~~bottom line for each object is the line fluxe in $\log$ erg cm$^{-2}$ s$^{-2}$}\\
\multicolumn{7}{l}{$^b$Blended with \ion{Na}{1} absorption.}\\
\multicolumn{7}{l}{$^c$Measured by summing the flux across the line.}\\
\end{tabular}}
\end{table}

\begin{table}
\caption{Spectral Properties}
\rotatebox{90}{
{\scriptsize \begin{tabular}{cccccccccccc}
\hline
Star & SpT & $A_V$ & $d$  &  $T_{eff}$ & $M_*$ & $\log L_*$ & $R_*$ & $BJ_{obs}$  & $BJ_{acc}$ & $\log L_{acc}$ & $\log \dot{M}$\\
     &     &  mag.  & pc  & K  & $M_\odot$ & $\log L_\odot$ & $R_\odot$ & & & $\log L_\odot$ & $\log \dot{M}$ yr$^{-1}$\\
\hline
\hline
TW Hya           & K7     & 0.0 & 56  & 4060  & 0.77  & -0.77   & 0.83      & $1.85\pm0.10$ & $3.3\pm0.5$ & -1.7  & -9.1 \\
Hen 3-600A       & M3     & 0.0 & 34  & 3415 & 0.37  &  -1.10  & 0.81     & $0.86\pm0.06$ & $3.2\pm1.2$ & -2.5& -9.6 \\
Hen 3-600B       & M3.5   & 0.0 & 34  & 3342 & 0.29  & -1.27  & 0.69      & $0.35\pm0.06$ & -- &  $<-4.1$ & $<-11.1$\\
UScoCTIO 112     & M5.5   & 0.0 & 145 & 3058  & 0.090 & -1.97 & 0.37        & $0.43\pm0.03$ & -- &  $<-5.4$ & $<-12.2$ \\
UScoCTIO 128 & M7 & 0.0 & 145 & 2880 & 0.060 & -2.38 & 0.26  & $2.8\pm0.2$ & $3.4\pm0.3$ & -4.6  & -11.4\\
DENIS-P J160603.9-205644 & M7.25 & 0.0 & 145 & 2838 & 0.050 & -2.03 & 0.41  &  $2.1\pm0.5$ & $3.4\pm1.5$  & -5.1 & -11.6\\
Oph J162225-240515A   & M7.25  & 0.0 & 145  & 2838 & 0.054 & -2.43  & 0.25 & $<0.4$      & -- & $<-6.6$ & $<-13.3$\\
2MASS J12073347-3932540 & M8.25  & 0.0 & 52.4 & 2632 & 0.035 & -2.72 & 0.21 & $3.7\pm0.4$ & $4.6\pm0.7$ & -5.3 & -11.9\\
SSSPM J1102-3431 & M8.5   & 0.0 & 55.2 & 2550 & 0.026 & -2.71 & 0.23     & $1.4\pm0.2$ & $2.9\pm0.9$ &-6.3 & -12.8\\ 
USco J160723.82−221102.0  & M8.5  & 0.3 & 145 & 2550 & 0.027 & -2.74& 0.22& --          &  --      & $<-4.8^a$   & $<-11.3$  \\ 
USco J160603.75-221930.0 & M8.75  & 0.0 & 145 &  2478 & 0.033 & -3.02 & 0.17& $3.8\pm0.6$ & $4.0\pm0.8$ &-5.6 & -12.3  \\
Oph J162225-240515B   & M8.75  & 0.0 & 145  &  2478 & 0.021 & -2.76  & 0.23 & $>1.9$      & $>2.2$ & -6.2   & $-12.6$ \\
\hline
\multicolumn{10}{l}{$^a$Based on upper limit on the \ion{He}{1} $\lambda5876$ lines}
\\
\end{tabular}}}
\end{table}

\clearpage

\begin{figure}
\epsscale{0.7}
\plotone{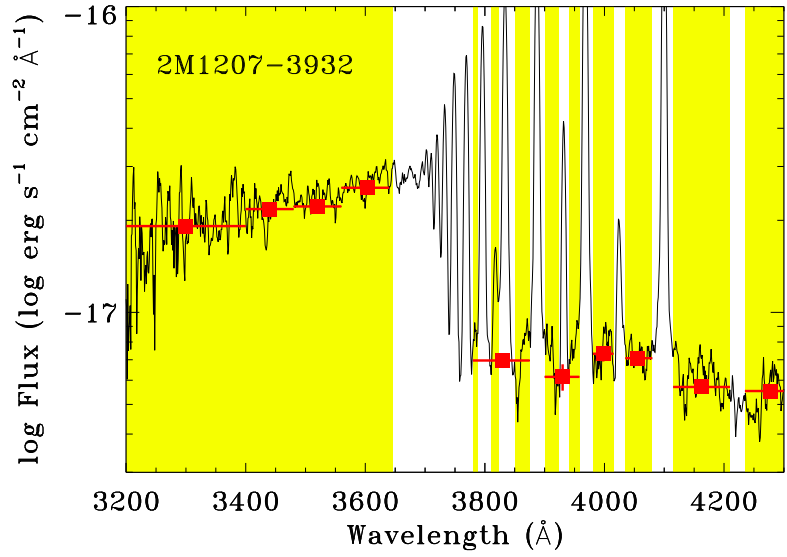}
\caption{The blue spectrum of 2MASS J12073347-3932540 shows strong excess continuum and line emission.  The square points are fluxes measured over continuum regions, which are shaded in yellow.  The bars in the x-direction describe the wavelength range covered by each point.  Although several of these ranges include emission lines in the unshaded regions, these regions are not included in our continuum flux esimate.  This method is applied to fainter targets to measure the continuum flux (Figure 4), which is required for our accretion rate estimates.}
\end{figure}

\begin{figure}
\epsscale{1.0}
\plotone{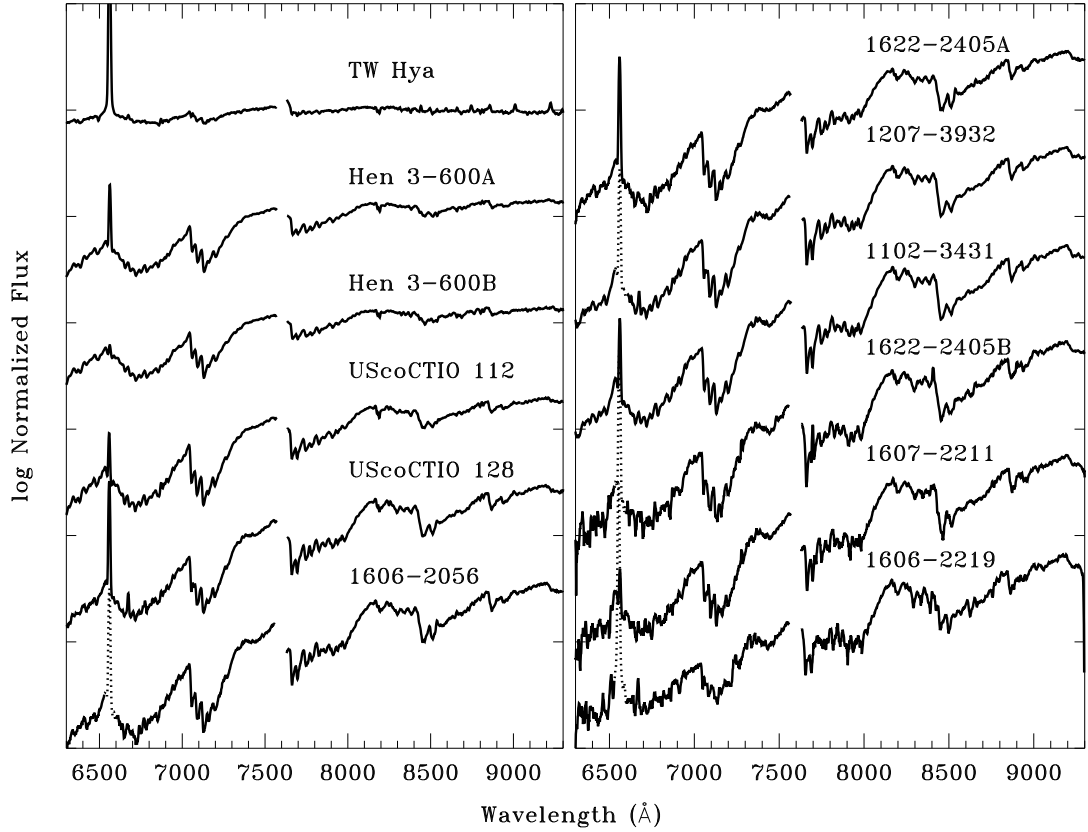}
\caption{Keck I/LRIS red spectra of nine brown dwarfs and three stars in USco and TWA, in order of earliest to latest spectral type in our sample.  The region of telluric absorption at 7600 \AA\ is not shown.  The H$\alpha$ emission line from several targets is plotted as a dotted line so that it does not cover the above spectrum.}
\end{figure}

\begin{figure}
\epsscale{0.7}
\plotone{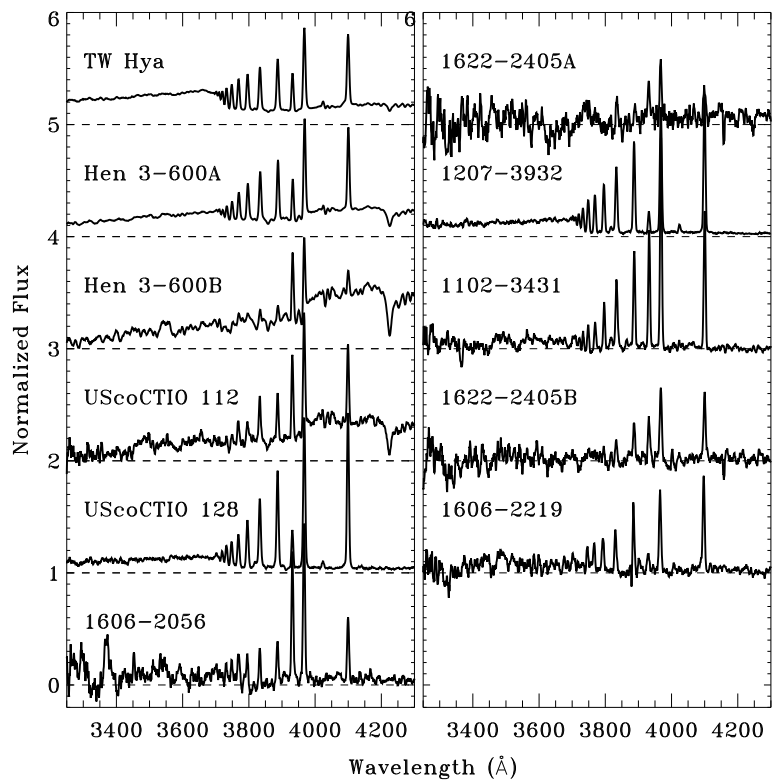}
\caption{{\it Keck}/LRIS blue spectra of three young stars and eight brown dwarfs, extracted at the full $R\sim700$ resolution of the data.  At this resolution the Balmer continuum is detected from TW Hya, Hen 3-600A, 2MASS J12073347-3932540, UScoCTIO 128, and undetected from Hen 3-600B.  For the fainter targets, the continuum emission at $<4500$ \AA\ is only detected after binning the continuum regions (shaded regions of Fig. 1).  Results from that low-resolution continuum extraction are shown in Fig. 4.  USco J160723.82−221102.0 is not shown here because no blue emission was detected from that target.}
\end{figure}

\begin{figure}
\epsscale{1.1}
\plottwo{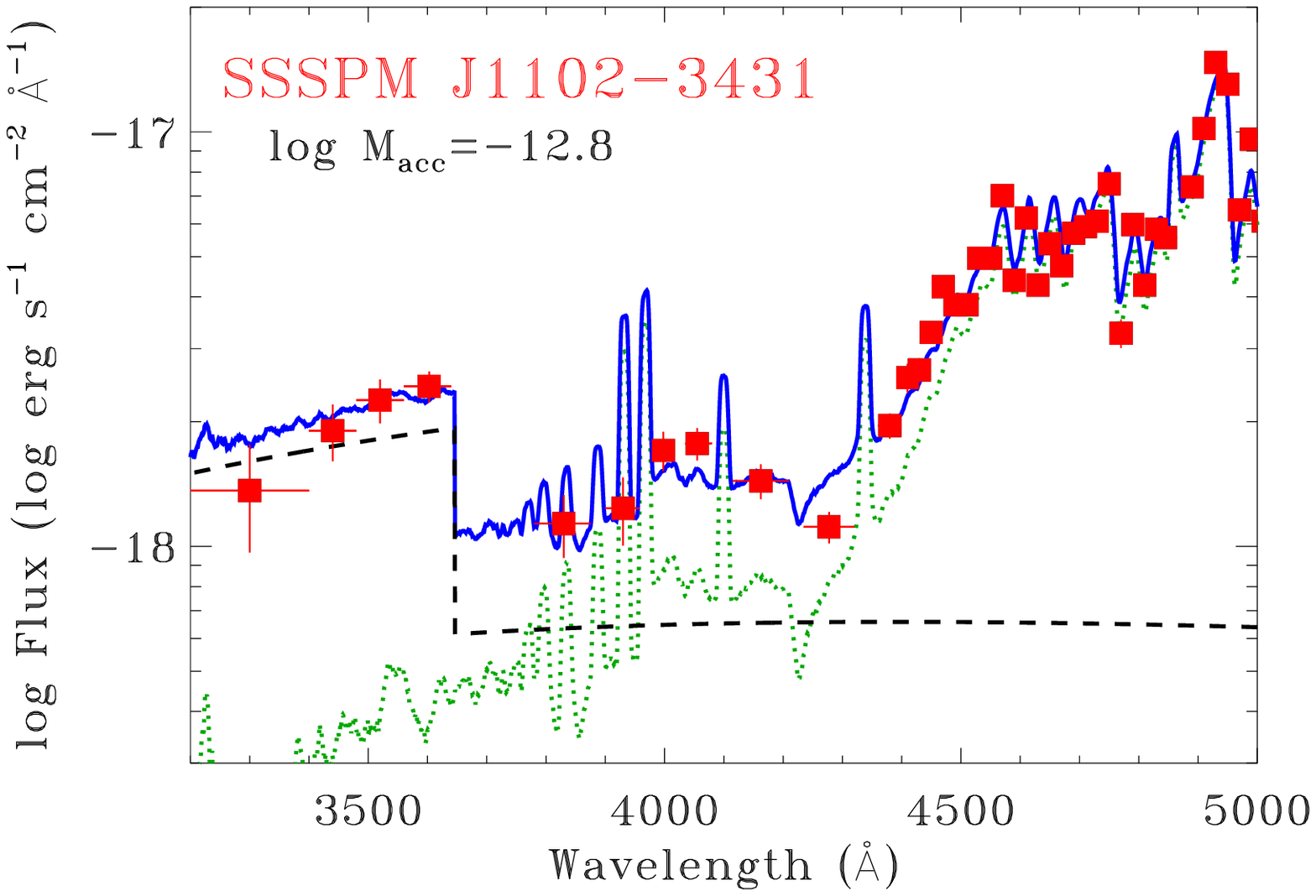}{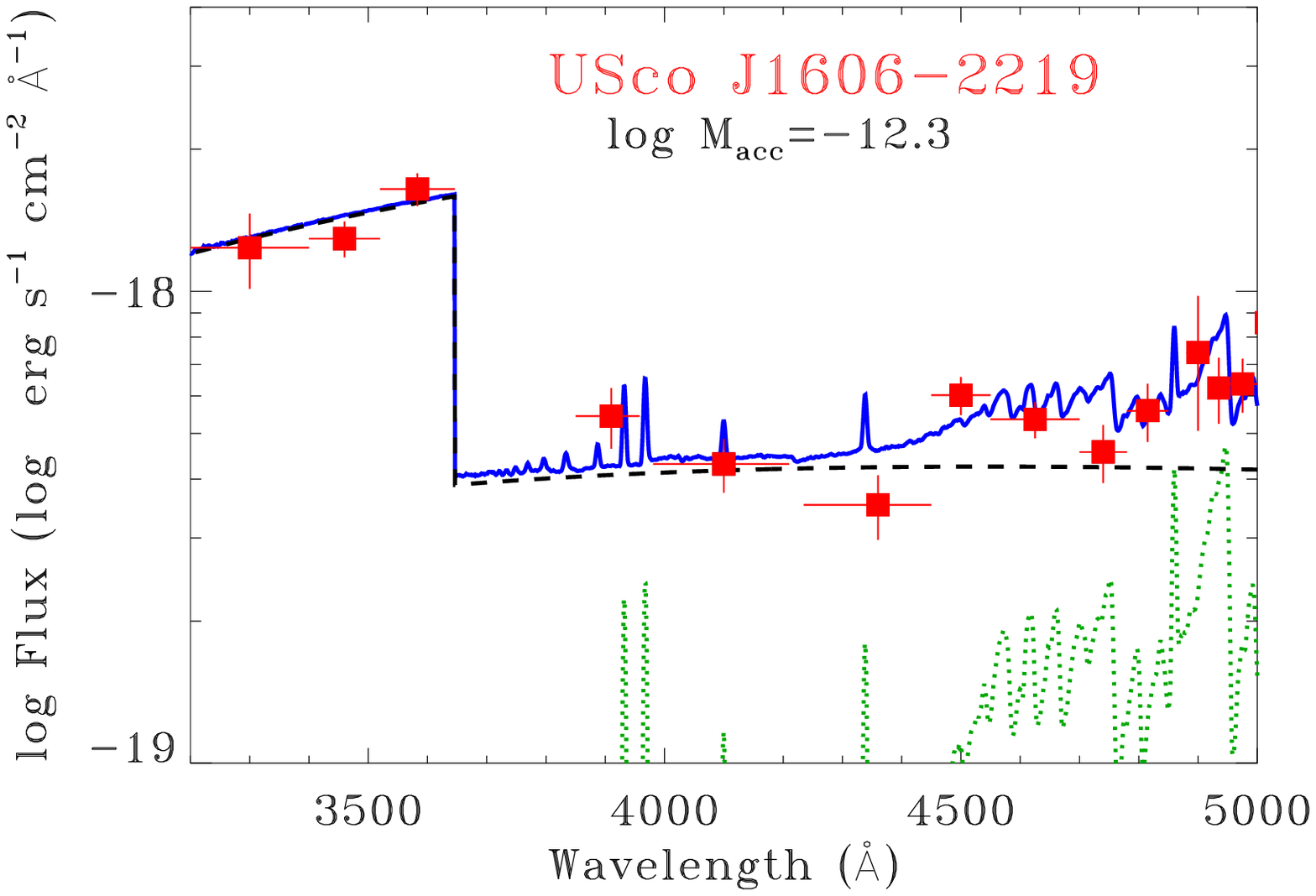}
\plottwo{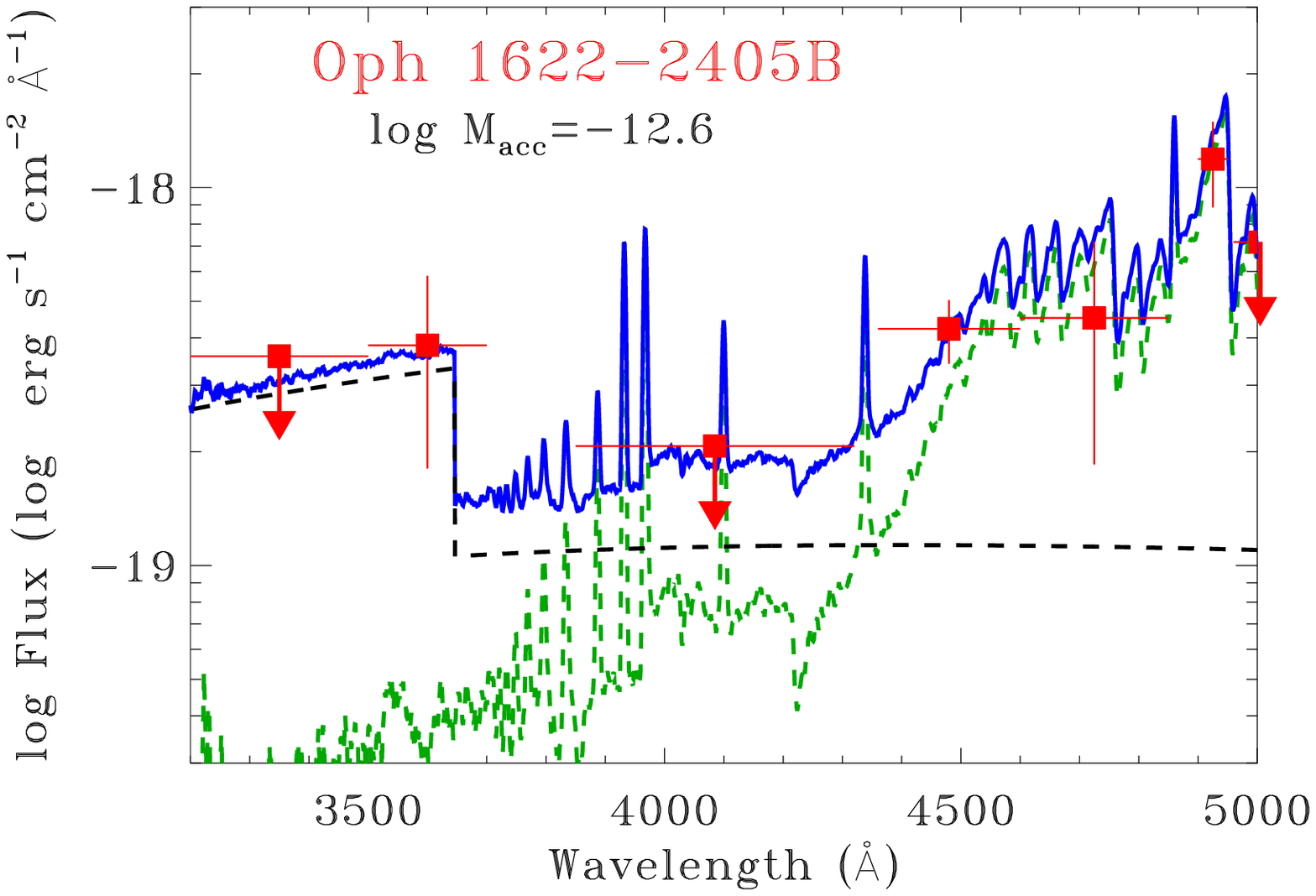}{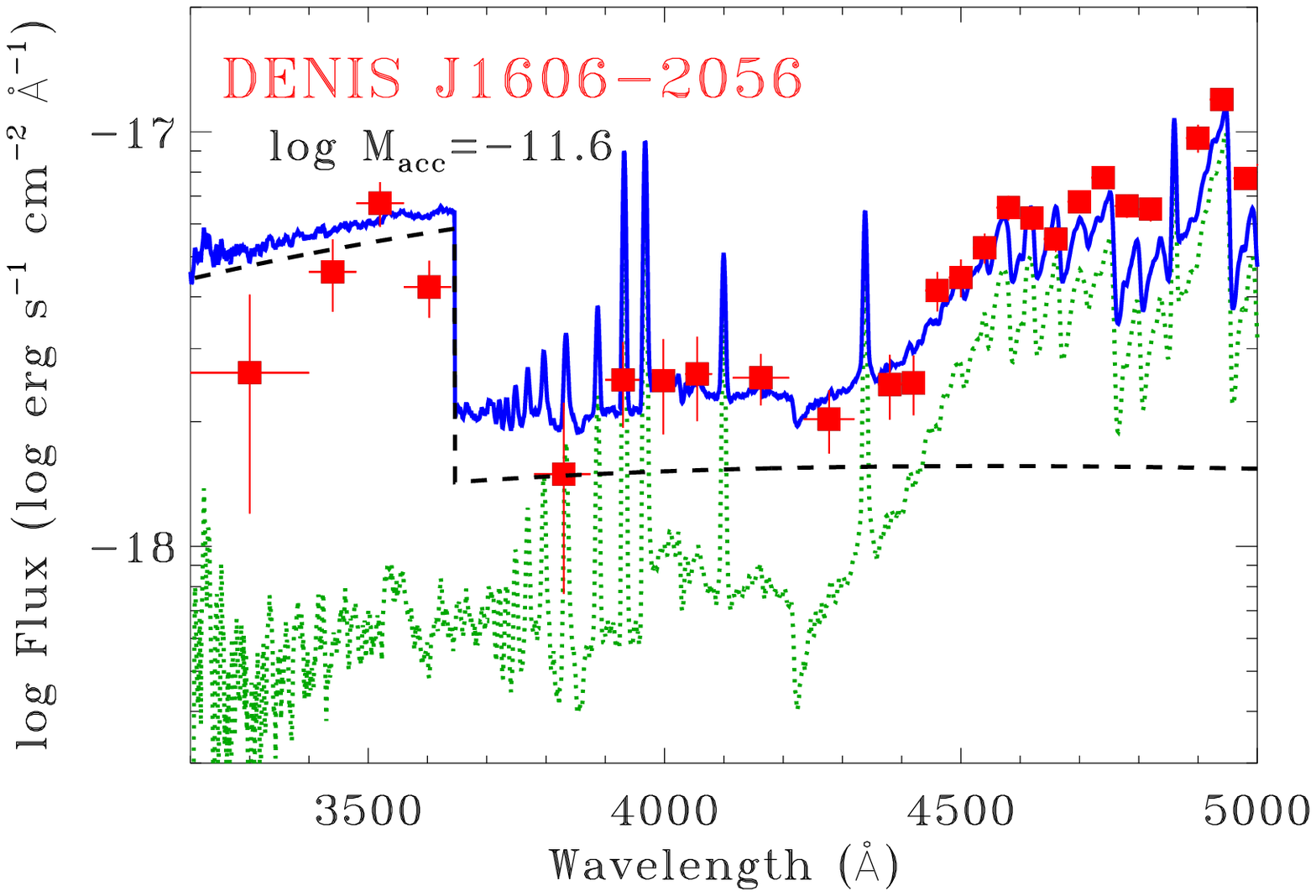}
\plottwo{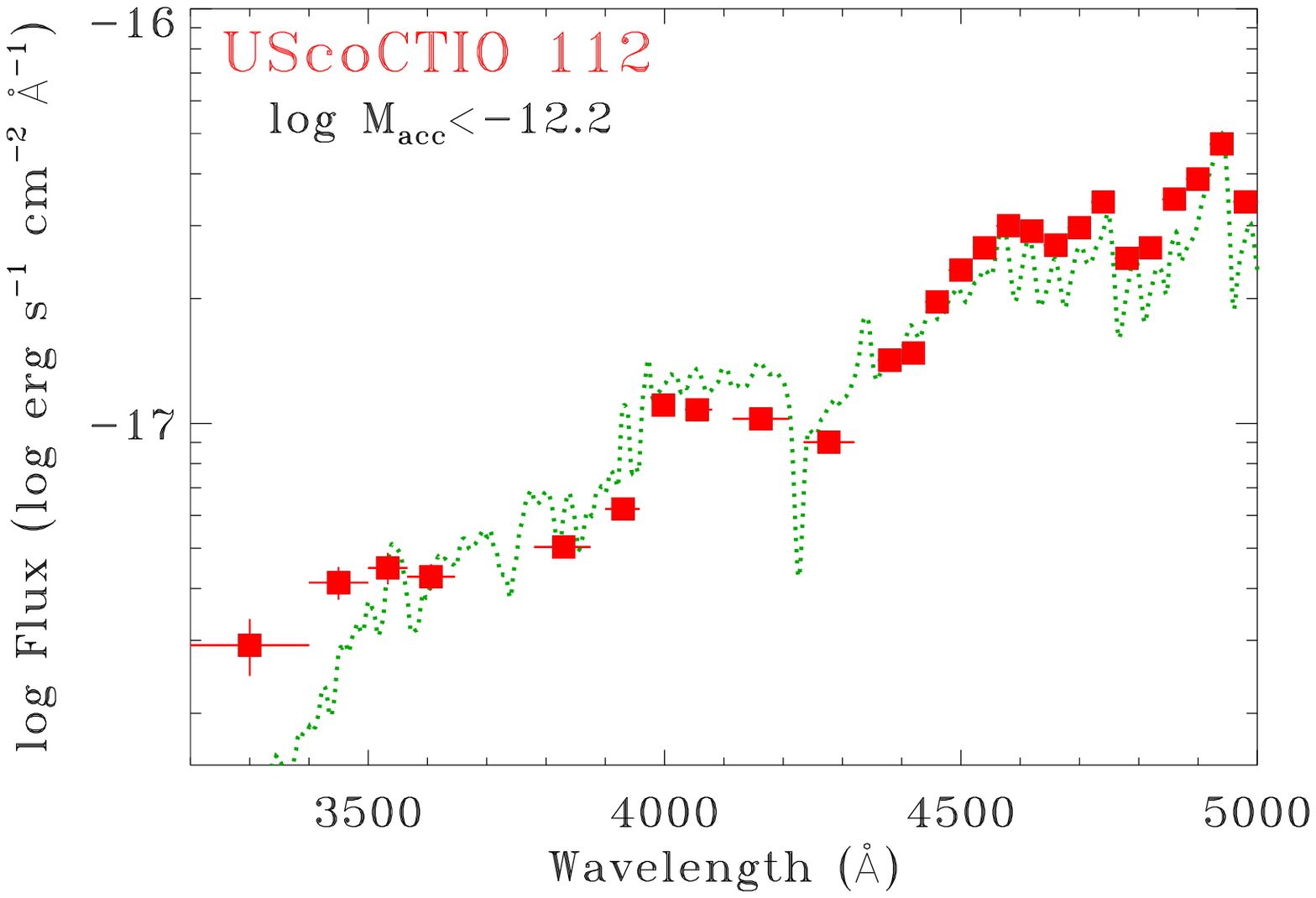}{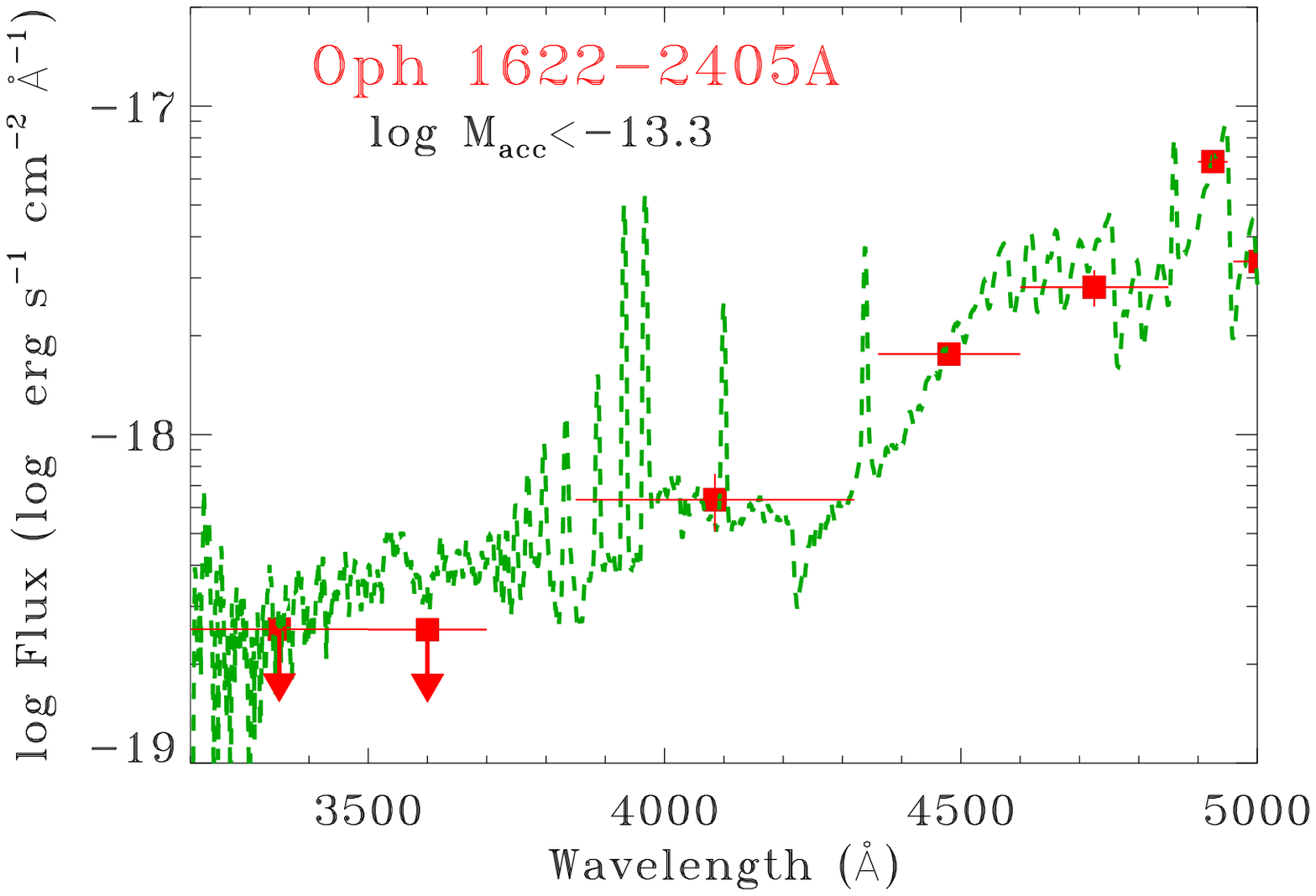}
\caption{Continuum flux measurements for six faint young brown dwarfs (red squares).  The dashed green line shows the spectrum of the photospheric template LHS 3003, scaled to the photospheric emission of the brown dwarf.  The solid blue line shows the spectrum of LHS3003 plus the accretion continuum, calculated as described in  \citet{Val93,Her08}.  Excess emission is detected from SSSPM J1102-3431, USco J160603.75-221930.0, DENIS-P J160603.9-205644, and Oph J162225-240515B but not from UScoCTIO 112 or Oph J162225-240515A.}
\end{figure}

\begin{figure}
\epsscale{0.6}
\plotone{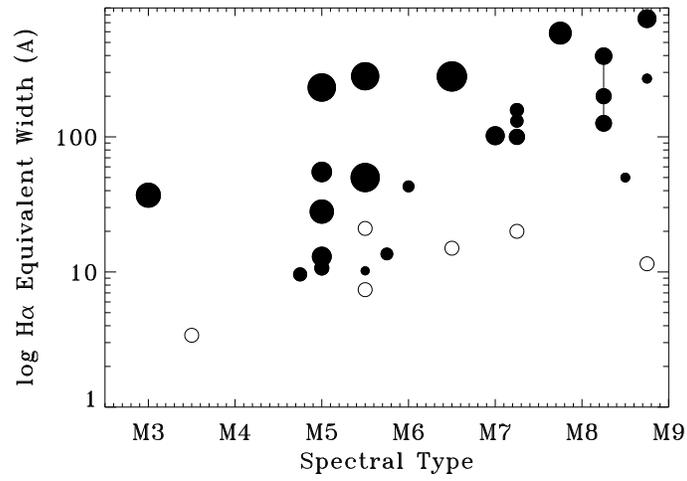}
\caption{H$\alpha$ equivalent width versus spectral type.  The size of the filled circles are determined by $L_{acc}/L_{bol}$.  Empty circles are objects that show no indication of excess Balmer continuum emission and are classified as non-accretors.  The line connects three observations of 2MASS J1207-3932.}
\end{figure}

\end{document}